\documentclass[preprint,12pt]{aastex}

\usepackage{graphicx} 
\usepackage{multicol}
\usepackage{natbib}
\usepackage{soul}

\newcommand{\be}{\begin{equation}}
\newcommand{\ee}{\end{equation}}
\newcommand{\nn}{\mbox{} \nonumber \\ \mbox{} }
\newcommand{\ba}{\begin{eqnarray}}
\newcommand{\ea}{\end{eqnarray}}

\newcommand\eg{\textit{e.g.}}
\newcommand\cf{\textit{cf.}}

\newcommand{\Bf}{{magnetic field}} 
\newcommand{\Bfs}{{magnetic field}}
\newcommand{\Ef}{{electric field}} 
\newcommand{\Efs}{{electric field}}

\newcommand{\E}{{\bf E}}
\newcommand{\B}{{\bf B}}

\newcommand{\J}{{\bf J}}

\begin{document}

\title{Coronal mass ejections as expanding force-free  structures}

\author{Maxim Lyutikov and Konstantinos N. Gourgouliatos}

\affil{
Department of Physics, Purdue University, 
 525 Northwestern Avenue,
West Lafayette, IN
47907-2036 }




\begin{abstract}
We model solar coronal mass ejections (CMEs) as expanding  force-free magnetic structures and find 
 the  self-similar  dynamics of   configurations with spatially constant $\alpha$, where  $\J =\alpha \B$,  in spherical and cylindrical geometries, expanding spheromaks and Lundquist fields respectively.    The field structures remain force-free, under the conventional non-relativistic assumption that the dynamical effects of the inductive electric fields can be neglected.  
 While keeping the internal magnetic field structure of the stationary solutions, expansion leads to complicated internal velocities and rotation, caused by inductive electric fields. The structure depends only on overall radius  $R(t)$ and rate of expansion $\dot{R}(t)$ measured at a given moment, and thus is applicable to arbitrary expansion laws.  
  In case of  cylindrical Lundquist fields, magnetic flux conservation requires  that both axial  and radial expansion proceed with  equal rates. In accordance with observations, the model predicts that  the maximum magnetic field is reached before the spacecraft reaches the geometric center of a CME. 
\end{abstract}

\maketitle

\section{Introduction}

Expansion of magnetic clouds is one of the basic problems in space physics and geophysics, related to propagation of  solar disturbances,  the coronal mass ejections  (CMEs) through interplanetary medium \citep{ForbesCME}.  
As the space craft passes through the cloud, the magnetic 
field	strength	is higher	than	average,	the density is lower,  the magnetic pressure inside the cloud  greatly exceeds the ion thermal pressure and the magnetic field direction changes through the cloud \citep[\eg][]{1982GeoRL...9.1317B}. 

There are two basic models of magnetic clouds: (i) magnetic flux ropes, still connected to the Sun at the distances of the Earth \citep{Burlaga81,Cargill95,CG1993}; (ii)  disconnected entities of spherical topology, which in the simplest paradigm can be modeled as spheromaks \citep[\eg][]{Ivanov1985,Vandas1991}. These are stationary solutions, while the CMEs expand while propagating in the Solar wind. 
Previous attempts to find a structure of expanding  spherical configurations have  led to breakdown of force-free condition through appearance of pressure and inertial effects \citep{Low82,Ivanov89,Vandas97}. 
In this paper  we find the structure of expanding magnetic configurations that {\it remain force-free}. In the force-free approach the dynamics of the system is dominated by the magnetic field and any plasma pressure or inertia terms are negligible, for that reason we shall not include in this study the equations of continuity of the fluid and we shall focus only on the magnetic field and on the induced electric field.

 The importance of spheromaks in spherical geometry \citep{ChandrasekharKendall57}   and  of \cite{Lundquist50}  fields  (in cylindrical geometry)  as examples of clouds with self-contained \Bfs\ stems from the fact that these solutions  are, in some sense, the minimal energy states and thus are stable to many ideal and resistive instabilities. 
\cite{Woltjer58} and \cite{1974PhRvL..33.1139T}  formulated the plasma relaxation principle, according to which plasma evolution conserves magnetic helicity and, if the magnetic field is the only important quantity, it evolves to a force-free state with constant $\alpha$, where ${\bf J} =\alpha {\bf B}$, where $\alpha$ has some similarities to thermodynamic temperature, so a relaxed state should have constant $\alpha$ \citep{Bellanbook}.

\section{Expanding Spheromaks}

\subsection{Magnetic Field of a Spheromak}

Spheromaks are stationary force-free configurations of plasma satisfying the condition ${\bf J} =\alpha {\bf B}$ with spatially constant $\alpha$.  They are solutions of  Grad-Shafranov equation in spherical coordinates when the poloidal current is a linear function of the magnetic flux. Below we will use ``the basic spheromak" solution, corresponding to axially symmetric  dipolar-like fields 
 \citep{ChandrasekharKendall57}.
 
Given  the magnetic flux function $P(r,\theta)$ and poloidal current function  $I(P)$, the \Bf\ is 
\be
{\bf B}= {\nabla P \over r  \sin \theta} \times {\bf e}_\phi+ {2 I (P) \over r  \sin \theta}  {\bf e}_\phi.
\label{B}
\ee
When only magnetic force is present, the magnetic flux function satisfies  the Grad-Shafranov equation
\be 
\Delta ^\ast P + 4 I I'=0,
\label{GS}
\ee
where
\be
\Delta ^\ast  = \partial_r^2 + {\sin \theta \over r^2}  \partial_\theta { \partial_\theta \over \sin \theta}
\ee
is the Grad-Shafranov operator.
For linear dependence $I = \alpha P/2$ with spatially constant  $\alpha$, Equation (\ref{GS}) has separable solutions in term of spherical Bessel functions $j_n ( \alpha r)$.
The first order solutions are then
\ba &&
P_0  =  B_0 {r \over \alpha} j_1 (\alpha r)  \sin^2 \theta,
\nn && 
B_r = 2 B_0 { j_1 \over \alpha r} \cos \theta,
\nn &&
B_\theta =-B_0 { j_1 + \alpha r j_1' \over \alpha r} \sin \theta,
\nn &&
B_\phi = B_0 j_1 \sin \theta.
\label{spheromak}
\ea
The first order  spherical Bessel function $j_1$  can be expressed in terms of elementary functions, $j_1(x) = \sin x /x^2 - \cos x /x$.
Parameter $\alpha$ is related to the size of a spheromak, defined by the surface where radial \Bf\ is zero,  solution of $j_1=0$, $R = C_\alpha/\alpha$, $C_\alpha =4.49$.

\subsection{Radial Expansion of a Spheromak}
\label{Ideal}

We now study the self-similar, non-relativistic expansion of axially symmetric  force-free structures with spatially constant $\alpha$-parameter.  Below we discuss a procedure that allows to {\it derive  non-relativistic, time-dependent force-free solutions from a given stationary force-free configuration}. First, we note that parameter $\alpha$ has  dimension of inverse length, and thus is related to overall size of the magnetic structure $R$. For expanding solutions we take  $\alpha=\alpha(t)$, but still spatially independent.
 According to self-similar prescription, we take all radial dependence to be in a form $  \propto r \alpha \propto r/R(t)$.
  The   induction equation, $\nabla \times \E + \partial_t \B =0$,  as well as  magnetic magnetic flux conservation, require that 
in a time-dependent case ${P}= (\alpha/\alpha_0)^2 P_0$, where $\alpha_0$ is the value at some initial time and $P_0$ is a solution of a stationary force-free problem (\eg\ given by Equation (\ref{spheromak})). Next, we are looking for ideal solutions, where \Ef\ is always perpendicular  
to \Bf\ and thus can be written as  $\E = -  {\bf V} \times \B$. Since we are looking for radially self-similar solutions, ${\bf V} =  V {\bf e}_r$ \cite[\cf][and Equation (\ref{v})]{Low82}. 
{This is the velocity of radial expansion. The drift velocity of the magnetic field lines is given by ${\bf v}= \E \times \B /B^{2}$ and is by definition perpendicular to the magnetic field lines. These two approaches give different values for ${\bf V}$ and ${\bf v}$. Using vector identities we can conclude that the difference between two velocities is a vector parallel to the magnetic field lines. Indeed} 
\begin{eqnarray}
{\bf v}=\frac{\E \times \B}{B^{2}}=-\frac{{\bf V} \times \B}{B^{2}}\times \B={\bf V}-\frac{{\bf V} \cdot \B}{B^{2}} \B\,.
\label{Velocities}
\end{eqnarray}
{There is a freedom of choice between these velocities without altering at all the force-free solution. Equilibria of this type, which contain also some matter however, do not allow transition from one velocity to the other without changing the dynamics of the problems just by adding a velocity component parallel to the magnetic field. In that case, an azimuthal component in the velocity will give rise to inertial terms in the force balance equation. In this study we focus in force-free equilibria where the dynamics of the system are dominated by the magnetic field  and inertia is negligible.} Choosing self-similar scaling, $V =  - r \partial_t \ln  \alpha$, we  find
 for axially symmetric \Bf, given by  Equation (\ref{B}),  the following  \Ef\
\be 
\E =  r { \dot{\alpha}  \over \alpha}  {\bf e}_r \times \B,
\ee
where dot denotes differentiation with time. 
It follows from Equation (\ref{Velocities}) that:
\be 
{\bf v}= { \B  (\B \cdot {\bf e}_r)- {\bf e}_r B^2 \over B^2} { r  \partial_t  \ln \alpha }.
\label{v}
\ee
In deriving the solution we neglected dynamical effects of \Efs, in particular displacement current and electric charge density, so that the force-free condition involves only \Bf\ and electric current, $\J \times \B=0$. This is indeed a reasonable assumption for the solar system where the velocities are non-relativistic.  

  For the basic spheromak solutions (\ref{spheromak}), the electromagnetic fields become
 \ba &&
 B_r =  2 B_0  \alpha { j_1 \over \alpha_0^2 r} \cos \theta, 
\nn &&
B_\theta =-B_0 \alpha  { j_1 + \alpha r j_1' \over \alpha_0^2 r} \sin \theta,
\nn &&
B_\phi = B_0 j_1 \left({\alpha\over \alpha_0}\right)^2  \sin \theta,
\nn &&
E_r=0,
\nn &&
E_\theta= - B_0  { \alpha \dot{\alpha} \over  \alpha_0^2}  r j_1 \sin \theta=- B_\phi {\dot{\alpha} \over \alpha} r,
\nn &&
E_\phi =- B_0  \dot{\alpha}  { j_1 + \alpha r j_1' \over \alpha_0^2 } \sin \theta= B_\theta {\dot{\alpha} \over \alpha} r,
\nn &&
{\bf E}=   \left( {\dot{\alpha} \over \alpha} r  \right) {\bf e}_{r} \times {\bf B}.
\label{main}
 \ea
We stress that the fields remain force-free, $\J \times \B=0$. 

  \subsection{Velocity Structure of the Solution}
 
\begin{figure}[h!]
   \includegraphics[width=0.99\linewidth]{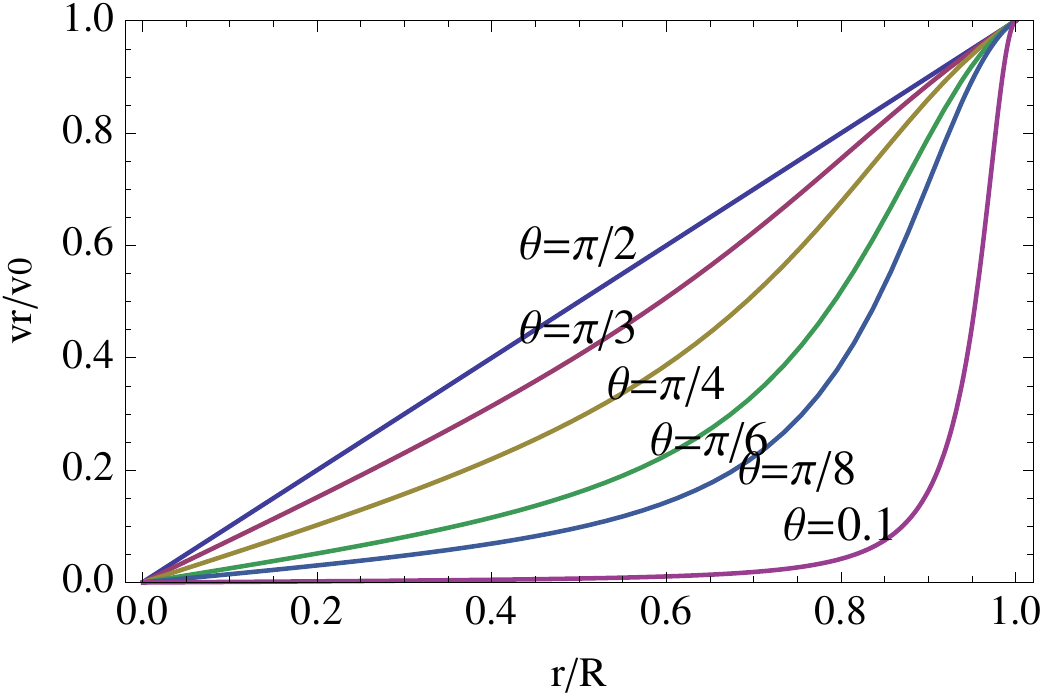}
   \caption{ Radial velocity $v_r$, $v_\theta$, $v_\phi$ as functions of $r $ for different $\theta$. In the equatorial plane radial velocity increases linearly  with radius, $v_r(\theta = \pi/2) = (r/R) v_0$, while close to the axis velocity remains small in the bulk, sharply increasing to $v_0$ near the surface.
   }
 \label{flowr}
 \end{figure}

\begin{figure}[h!]
         \includegraphics[width=0.99\linewidth]{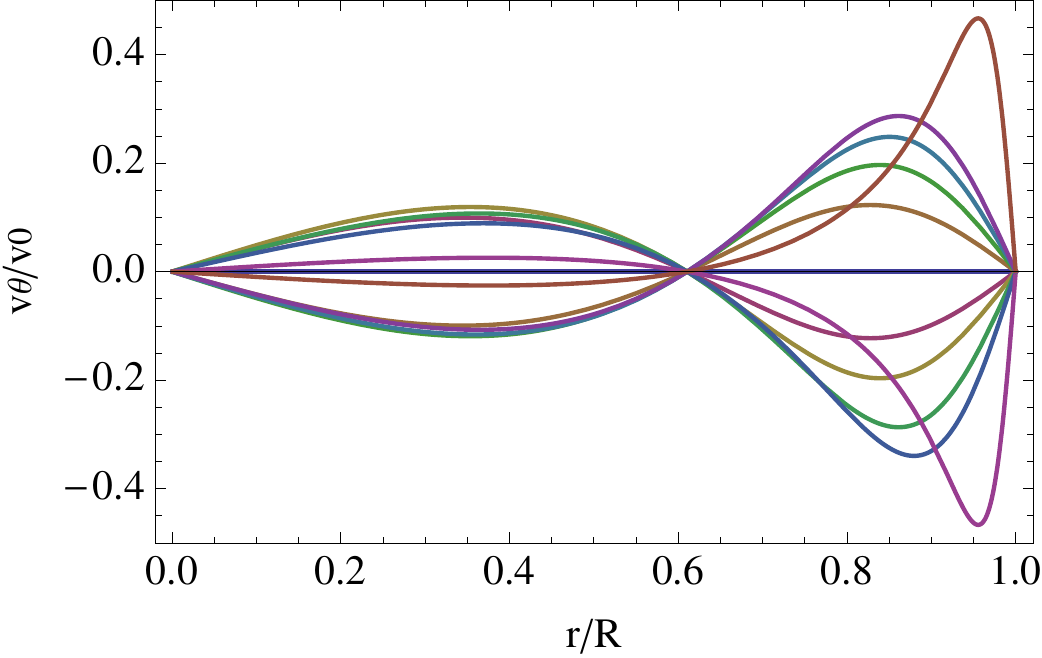}
   \caption{ Velocity $v_\theta$ as a  functions of $r$ for different $\theta$.  Flows are antisymmetric with respect to the equatorial plane.  $v_\theta$ becomes zero at $r=0.61 R$ (root of the equation 
 $E_\phi=0$)}
    \label{flowtheta}
 \end{figure}

\begin{figure}[h!]
          \includegraphics[width=0.99\linewidth]{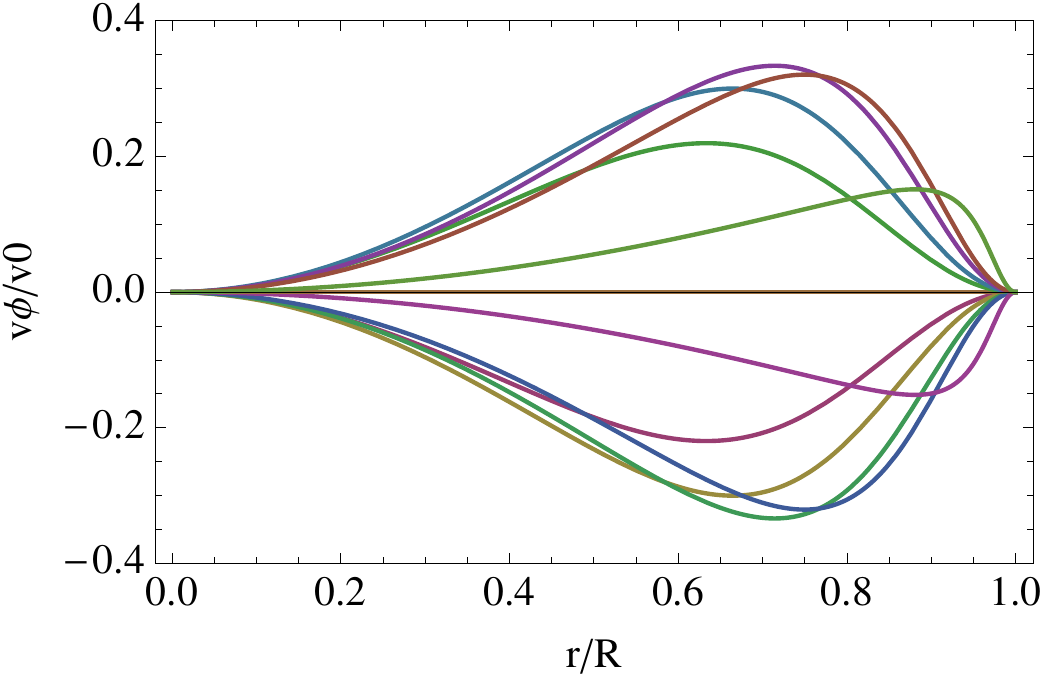}
   \caption{ Velocities  $v_\phi$ as functions of  $r$ for different $\theta$. Flows are antisymmetric with respect to the equatorial plane.
   }
 \label{flowphi}
 \end{figure}

 Expanding force-free spheromaks have complicated internal velocity structure. 
 At each point there is a well defined electromagnetic velocity 
 ${\bf v} =   {\bf E} \times {\bf B} / B^2$, normalised to $c$, Equation (\ref{v}).
 The solution can be  parametrized by the velocity of the expansion of the boundary, which is
 $v_\theta=v_\phi=0$ and $v_r=  R(t) \partial_t \ln  (1/\alpha) =v_0$, see Figures \ref{flowr},\ref{flowtheta},\ref{flowphi}.
 
\begin{figure}[h!]
   \includegraphics[width=0.49\linewidth]{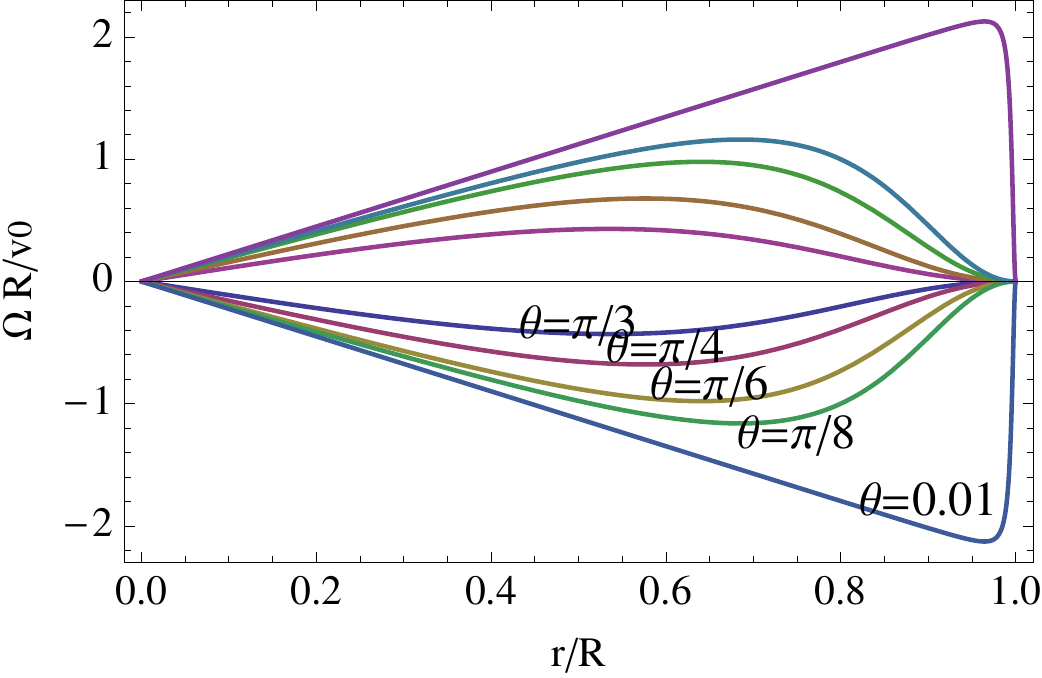}
      \includegraphics[width=0.49\linewidth]{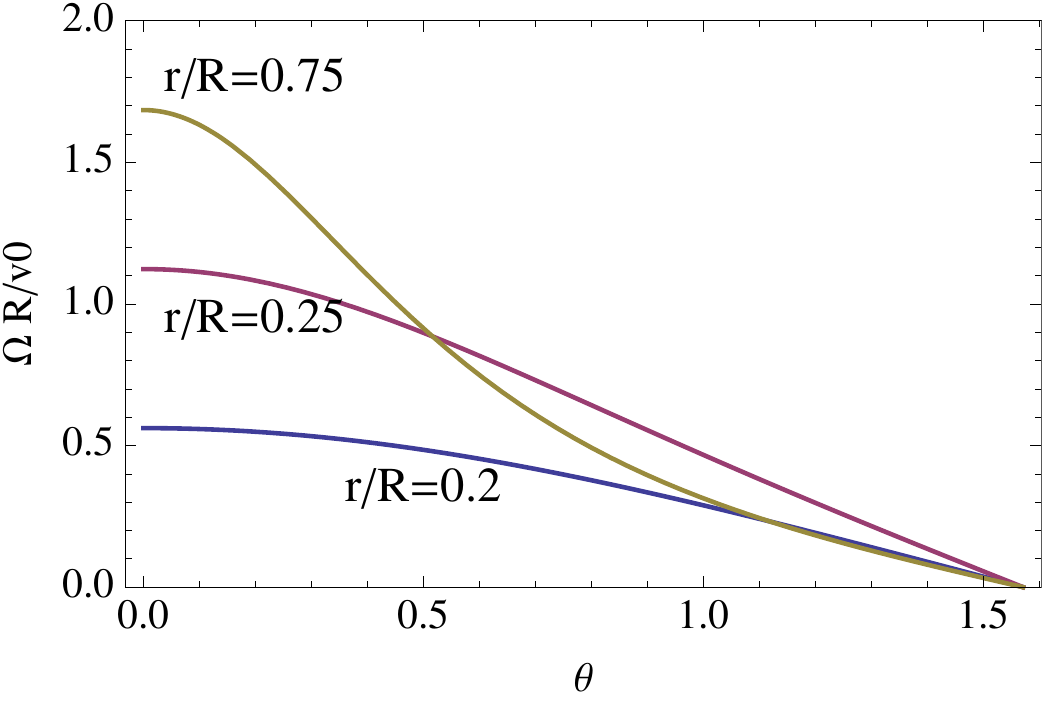}
   \caption{ Angular velocity of rotation $\Omega= v_\phi/( r \sin \theta)$  as function of radius and polar angle.
   }
 \label{omega}
 \end{figure}

\begin{figure}[h!]
   \includegraphics[width=0.99\linewidth]{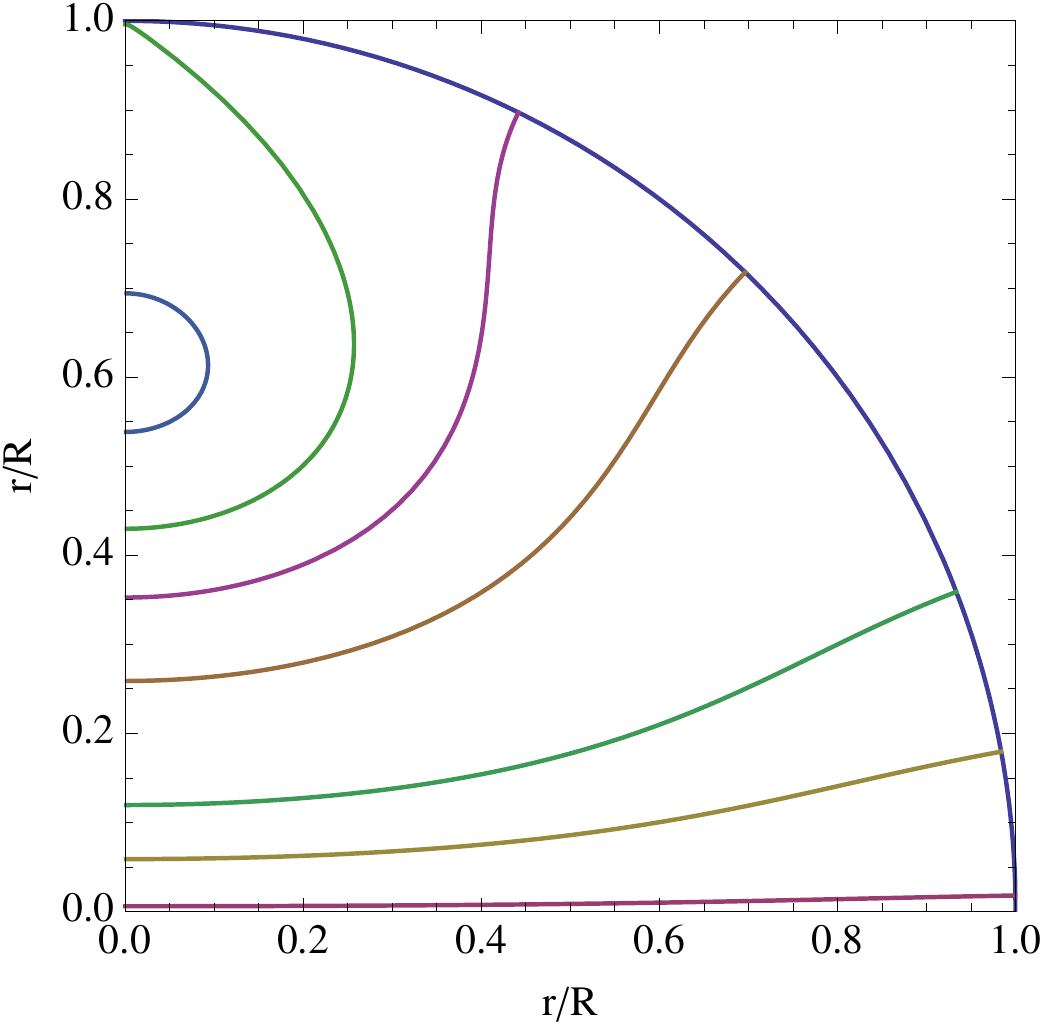}
   \caption{ Plasma flow lines. On the surface flow lines become radial to match the overall   radial expansion.   The stationary point is 
   $\theta=0$, $r=0.61 R$ (root of the equation 
 $E_\phi=0$).   }
 \label{flow1}
 \end{figure}

Since $v_\phi \neq 0$, the expansion of a spheromak induces  rotation, with the different hemispheres rotating in the opposite direction  (Figure \ref{omega}).
The maximum toroidal velocity is $0.33 v_0$ reached at $\theta=0.53$ and $r/R =0.71$.
For $r\ll R$ the angular velocity of rotation is  $\omega \sim \pm {  C_\alpha v_0 \sin 2 \theta} { r\over 4 R^2}$.
 The flow lines are plotted in Figure \ref{flow1}.
 Close to the symmetry axis, velocity at small radii is directed away from the axis.. 
Total velocity at small $r \ll R$ is $v  \sim v_0 {r \sin \theta /R}$. The flow lines are orthogonal to the surface, matching the overall expansion of the spheromak.

Using expressions (\ref{main}) for electromagnetic fields, we can calculate the  total energy of the expanding spheromak
\be
E_{\rm tot} = \int d V { E^2 + B^2 \over 8 \pi} =  \left( c_1 + c_2 {v^2 \over c^2} \right) { B_0^2 R_0^4 \over R}
\ee
where $c_1= 2.5 \times 10^{-3}$ and $c_2=8.7 \times 10^{-4}$ are combinations of power laws and trigonometric functions of $C_\alpha$. 
The second term proportional to velocity squared is the contribution of the \Ef\ and can be ignored for non-relativistic expansion. 

It may be verified that the total magnetic helicity, $ {\cal H} = \int dV {\bf B} \cdot {\bf A}=1.4 \times 10^{-2} B_0^2 R_ 0^4$, as well as  the toroidal magnetic flux 
${\cal F} =   \int  B_\phi r dr= 5.26 B_0 R_0^2$  are  independent of time. At the same time, the total electromagnetic energy decreases during expansion $\propto 1/R$. This highlights an important point, that  it is impossible to conserve both electromagnetic  energy and magnetic helicity. In case of ideal expansion the energy decrease is due to $pdV$ work done on the surrounding plasma. Magnetic pressure on the surface  $\propto 1/R^4$, energy  $\propto 1/R$.

We can compare the self-similar spheromak solutions to the  internal energy of adiabatically expanding polytropic gas with $p \propto \rho^\gamma$,  
$E \propto p V \propto R^{3 (1-\gamma)}$. Thus, effectively, spheromak behaves as a  polytropic gas with $\gamma=4/3$.  This is related to the results of  \cite{Low82}, who found self-similar MHD solutions for expanding cloud if the fluid obeys polytropic law with $\gamma=4/3$. In this case both magnetic and thermal components of plasma behave similarly \citep{GV2010}. 

\section{Toroidal Configurations}

In the case where magnetic clouds are connected to the Sun \citep{CG1993}, their topology shall resemble more an arcade, and their mathematical description is more accurately approximated by a force-free magnetic field confined in a torus. A number of studies \citep{Burlaga81, B1988, L1990} have considered solutions in cylindrical geometry as an approximation to the toroidal structures  of  magnetic cloud. 

\begin{figure}[h!]
  \includegraphics[width=0.4\linewidth]{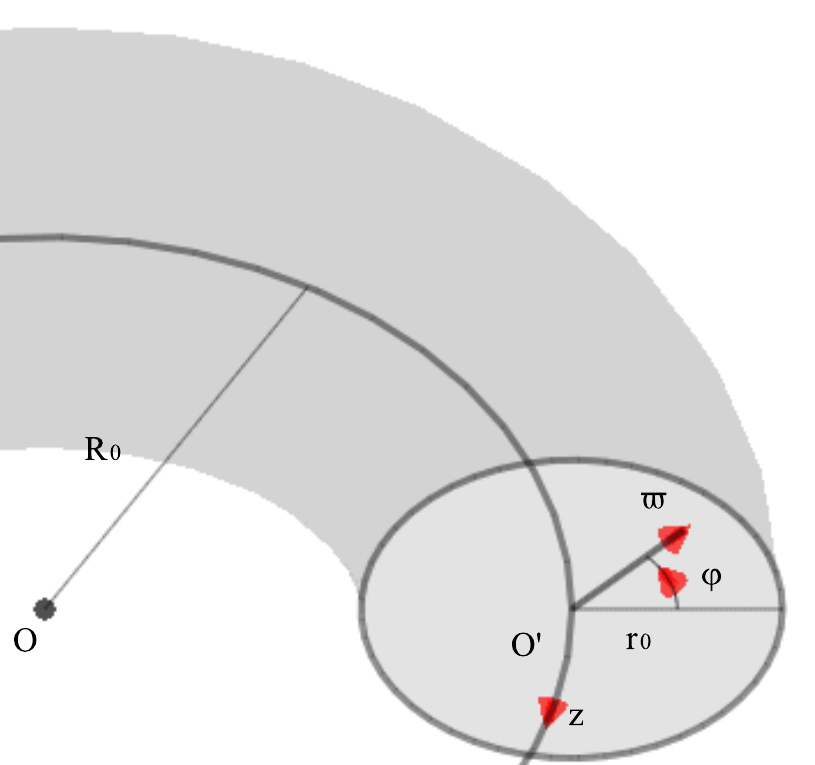}
  \caption{The geometry of the torus, the large radius $R_{0}$ writes a circle with centre $O$ and the small one $r_{0}$ a circle with centre $O'$. We attach the orthogonal coordinate system $(\varpi, \phi, z)$ to the circle $(O', r_{0})$.
  }
\label{torus}
\end{figure}

In this section we examine the effect of expansion in this type of solution by applying self-similarity. Let us assume a torus (Figure \ref{torus}) of large radius $R_{0}$ and small $r_{0}$, the centre corresponding to the large radius is $O$ and the centre corresponding to the small radius lies on the circle $(O, R_{0})$ and shall be called $O'$. Let us attach an orthogonal system of coordinates to $O'$; this coordinate system resembles a cylindrical coordinate system with the difference that the role of the axis $z$ is taken by the circle $(O, R_{0})$, the coordinates in this system are $(\varpi, \phi, z)$, where $0 \le \varpi \le r_{0}$, $0 \le \phi \le 2 \pi$ and $-\pi R_{0} \le z \le \pi R_{0}$; the field inside the torus has components $B_{\varpi}$, $B_{\phi}$ and $B_{z}$. Such fields have been studied by \cite{MT1981}, who considered force-free structures in toroidal coordinates and  found that for large aspect ratios $R_0/r_0\gg 1 $ the solution reduces to \cite{Lundquist50} fields.
 The first few terms of the Miller-Turner fields are
\begin{eqnarray}
B_{\varpi, {\rm T}}=\frac{B_{0}}{R_{0}\alpha}\Big(-J_{0}(\alpha \varpi)+\frac{r_{0} F}{\varpi}\Big)\sin\theta\,, \nonumber \\
B_{\phi, {\rm T}}=B_{0}\Big[J_{1}(\alpha \varpi)-\frac{1}{\alpha R_{0}}\Big(J_{0}(\alpha \varpi)-r_{0}\frac{d F}{d\varpi}\Big)\cos \theta\Big]\,, \nonumber \\
B_{z,{\rm T}}=B_{0}\Big(J_{0}(\alpha \varpi)-\frac{r_{0}}{R_{0}}F\cos\theta\Big)\,, \nonumber \\
F=\frac{\varpi}{2 r_{0}}J_{0}(\alpha \varpi)+\frac{1}{2}\frac{J_{0}(\alpha r_{0})}{J_{1}(\alpha r_{0})}J_{1}(\alpha \varpi) \,.
\end{eqnarray}
{Let us now consider a thin section of the torus which can be approximated with a cylinder.
In the large aspect ratios  limit, $R_0/r_0\gg 1 $  we only keep the Lundquist field.}
\begin{eqnarray}
B_{\varpi,{\rm S}}=0\,, \nonumber \\
B_{\phi,{\rm S}}=B_{0}J_{1}\Big({\alpha} \varpi\Big)\,, \nonumber \\
B_{z,{\rm S}}=B_{0}J_{0}\Big({\alpha} \varpi\Big)\,,
\label{B_STATIC}
\end{eqnarray}
where $J_{0,1}$ is the Bessel function of the zeroth and first order. 

Magnetic clouds expand as they propagate away from the Sun. The simultaneous expansion and propagation lead to an increase of both radii of the torus. The expansions of the small and large radii need to scale with the same factor to conserve the magnetic flux that crosses  the equator of the torus and the magnetic flux that crosses a section of the torus: $\Phi_{1}\propto B_{\phi,{\rm T}} r_{0}R_{0}$ and $\Phi_{2}\propto B_{z,{\rm T}} r_{0}^2$, so if $R_{0}$ is multiplied with some factor so does $r_{0}$. 

This field shall experience the effect of expansion both in $\varpi$ and $z$ directions. Let us consider an expansion velocity ${\bf V}=-\varpi\frac{\dot{\alpha}}{\alpha}{\bf \hat{e}}_{\varpi}-z\frac{\dot{\alpha}}{\alpha}{\bf \hat{e}}_{z}$.

Similar to Section \ref{Ideal}, the magnetic flux conservation  of the expanding structure requires that
\begin{eqnarray}
B_{\varpi,{\rm E}}=0\,, \nonumber \\
B_{\phi,{\rm E}}=B_{0}\Big(\frac{{\alpha}}{\alpha_{0}}\Big)^{2}J_{1}\Big({\alpha} \varpi\Big)\,, \nonumber \\
B_{z,{\rm E}}=B_{0}\Big(\frac{{\alpha}}{\alpha_{0}}\Big)^{2}J_{0}\Big({\alpha} \varpi\Big)\,.
\label{B_Expanding}
\end{eqnarray}
Then the electric field is $\E =- {\bf V} \times \B$, which gives
\begin{eqnarray}
E_{\varpi}=-B_{0}\frac{\dot{\alpha}{\alpha}}{\alpha_{0}^{2}}J_{1}\Big({\alpha}\varpi\Big)z\,, \nonumber \\
E_{\phi}=-B_{0}\frac{\dot{\alpha}{\alpha}}{\alpha_{0}^{2}}J_{0}\Big({\alpha}\varpi\Big)\varpi\,, \nonumber \\
E_{z}=B_{0}\frac{\dot{\alpha}{\alpha}}{\alpha_{0}^{2}}J_{1}\Big({\alpha}\varpi\Big)\varpi\,.
\end{eqnarray}
The whole structure is force-free satisfying $(\nabla \times \B) \times \B=0$, indeed
\begin{eqnarray}
\nabla \times \B=B_{0}\frac{\alpha^{3}}{\alpha_{0}^{2}}J_{1}(\alpha \varpi) {\bf \hat{e}_\phi} +B_{0}\frac{\alpha^{3}}{\alpha_{0}^{2}}J_{0}(\alpha \varpi){\bf \hat{e}_z}=\alpha \B\,.
\end{eqnarray}
In addition, the divergence of the magnetic field is zero $\nabla \cdot {\bf B}=0$ and the magnetic flux is conserved as it satisfies Faraday's law of induction $\nabla \times {\bf E}+\partial_{t} {\bf B}$=0, indeed
\begin{eqnarray}
\nabla \times \E=-B_{0}\frac{\dot{\alpha}\alpha}{\alpha_{0}}[(J_{1}(\alpha\varpi)+\alpha \varpi J_{0}(\alpha \varpi)){\bf \hat{e}_\phi}-(\alpha \varpi J_{1}(\alpha \varpi)-2J_{0}(\alpha \varpi)){\bf \hat{e}_z}]=-\partial_{t}\B \,.
\end{eqnarray}

We stress that though we focus in on a thin slice of the torus area near $z=0$, where the $V_{z}$ component vanishes,  we do take into account the fact that both the small and the large radii of the torus are increasing. The latter gives an expansion in the $z$ direction which creates a minor electric field in the $\varpi$ direction which is necessary  to conserve the magnetic flux. 

Again we evaluate drift velocity of the field lines ${\bf v}=\E \times \B/B^{2}$, whose components are
\begin{eqnarray}
v_{\varpi}=-\frac{\dot{\alpha}}{{\alpha}}\varpi \,, \nonumber \\
v_{\phi}=\frac{\dot{\alpha}}{{\alpha}}\frac{J_{0}\Big({\alpha}\varpi\Big)J_{1}\Big({\alpha}\varpi\Big)}{J_{0}\Big({\alpha}\varpi\Big)^{2}+J_{1}\Big({\alpha}\varpi\Big)^{2}}z \,, \nonumber \\
v_{z}=-\frac{\dot{\alpha}}{{\alpha}}\frac{J_{1}\Big({\alpha}\varpi\Big)^{2}}{J_{0}\Big({\alpha}\varpi\Big)^{2}+J_{1}\Big({\alpha}\varpi\Big)^{2}}z\,. 
\end{eqnarray}
The components of ${\bf v}$ shall match ${\bf V}$ at the boundary. Indeed, the $v_\varpi$ component is mathced, then from the $\phi$ component we find that the cylinder shall end either at a root of $J_{1}$ or $J_{0}$ and then from the $z$ component we find the acceptable solution is that of a root of $J_{0}$, thus the end of the rope is defined by the position where $\alpha \varpi$ equals a root of $J_{0}$. 

Formally, the $\phi$ and $z$ velocity components are proportional to $z$ and this breaks the translational symmetry of the Lundquist field. This is the outcome of choosing an arbitrary cross section and setting the velocity profile proportional to the distance from there. The physical system actually consists of a torus and as a result of the expansion an observer on a given point in the torus will see the other points to move away; and their distance measured on the $z$ coordinate will increase proportional to $z$. Given the arbitrary choice of the point $z=0$ we have the freedom to set this point at the place where the detector crosses the magnetic cloud, provided that its trajectory is perpendicular to the axis.  The detector will see a structure where the values are given for $z=0$, the magnetic field is that of Equation (\ref{B_Expanding}), the electric field though is
\begin{eqnarray}
E_{\varpi,0}=0\,, \nonumber \\
E_{\phi, 0}=-B_{0}\frac{\dot{\alpha}{\alpha}}{\alpha_{0}^{2}}J_{0}\Big({\alpha}\varpi\Big)\varpi\,, \nonumber \\
E_{z,0}=B_{0}\frac{\dot{\alpha}{\alpha}}{\alpha_{0}^{2}}J_{1}\Big({\alpha}\varpi\Big)\varpi\,.
\end{eqnarray}
Finally the velocity of the fluid is
\begin{eqnarray}
v_{\varpi , 0}=-\frac{\dot{\alpha}}{{\alpha}}\varpi\,, \nonumber \\
v_{\phi ,0}=v_{z,0}=0\,.
\end{eqnarray}
We remark that similar results for the cylindrical Lundquist type fields have been obtained recently and independently by \cite{Dalakishvili2010}.

\section{Matching to Driven Shock Solutions}

The magnetic clouds expand while propagating away from the Sun \citep{KB1982}. Next we consider how the expansion rate scales with the distance from the Sun. This depends both on the structure of the Solar wind  and on the conservation or not of the magnetic flux and helicity of the cloud, during its evolution, which in turn is related to the connection or not of the cloud to the Sun. A spheromak-type cloud is a disconnected entity which ideally conserves its magnetic flux and helicity, however as it expands its electromagnetic energy cannot remain constant. Magnetic flux tubes, however, are  attached to the Sun, so injection of magnetic flux or helicity is more likely.

 Let the spheromak have magnetic flux $\Phi$ enclosed within a volume of size $R$; the magnetic flux has dimensions $M^{1/2}L^{3/2}T^{-1}$.
 The magnetic field on the surface is not constant, but for simplicity we assume that it is of order $B_{0}=\Phi/R^{2}$. Let the external medium have density $\rho_{\rm ext}$,  which has dimensions $ML^{-3}$; the other variable is time whose dimension obviously is $T$. From dimensional analysis we find that expansion of a spheromak obeys a law: 
\be
R \sim \rho_{\rm ext}^{-1/6}\Phi^{1/3}t^{1/3}\,.
\label{R(t)}
\ee
A somewhat different approach leading to the same result comes  from balancing ram pressure of a strong forward shock, $\rho_{\rm ext} (dR/dt)^2$, with  \Bf\ energy density on the surface $  B^2 \sim(\Phi/R^{2})^{2}$.

For constant mangetic flux $\Phi$,  and assuming that external density scales with radius as  $\rho_{\rm ext}  \propto R^{-a}$, 
the self-similar expansion requires
 \be
R \propto t^{2/(6-a)}
 \label{point}
\ee
In a constant density medium this gives $R\propto t^{1/3}$. 
Note, that for self-similar scaling the energy in shocked plasma, $\propto R^2 \rho_{\rm ext}  \dot{R}^2 \propto t^{- 2/(6-a)} $ {\it decreases } with time, 
as $t^{-1/3}$ in constant density,  \citep[\cf][Equation (44)]{Farrugia95}. This can be achieved, \eg, if the shocked plasma cools radiatively, but the 
 energy loss should be tuned to the self-similar dynamics, which is not justified. 

Thus, generally,  if  a magnetic  
bubble drives  a non-radiative shock, the surface pressure inside the bubble falls off faster with time than the post-shock pressure and the self-similar assumption breaks down. At every moment the size of the bubble would adjust to the post-shock pressure in a non-self-similar way. 
For example, if initially the bubble is over pressurized with respect to a surrounding medium of constant density, expansion will drive a shock with 
$r_{\rm s} \propto t^{2/5}$  \citep{Sedov}.  The post-shock pressure scales as $p\propto t^{-6/5}$ and the size of  the magnetic  bubble $R\propto t^{3/10}$.  Thus, the whole system of magnetic bubble plus external shock medium evolves in a non-self-similar manner, with the relative size of the magnetic bubble decreasing $R/r_{\rm s}\propto t^{-1/10}$. 
Most energy is transferred to the outside plasma at initial stages of expansion.

\subsection{Expansion in a Wind Environment}

Motion of a CME in a wind is composed of translational motion  and overall expansion.
The translational velocities of the clouds vary  in the range of $300$km s$^{-1}$ to $1000$km s$^{-1}$ with average velocity $478$km s$^{-1}$ \citep{BS1998}. This velocity is of  the same order of magnitude as the Solar wind velocity.  \footnote{ We remark that if we take into account the buoyancy force and the drag due to the motion relative the solar wind we find that the limiting velocity is of the order of the sound speed of the solar wind, which is  a few $10^2$ km s$^{-1}$ depending on the temperature which is consistent with the numbers mentioned above. }  The  local density at  the cloud's location is then $\rho_{\rm ext} =   \rho_0 r_0^2/r_{\rm CME} ^2$, where   $r_{\rm CME} \sim v_{\rm SW}t$.   Substituting in Equation (\ref{R(t)}) we find that
\ba &&
R \sim \rho_{0}^{-1/6}r_{0}^{-1/3} v_{\rm SW}^{1/3}\Phi^{1/3}t^{2/3}
\nn &&
R \sim \rho_{0}^{-1/6} r_{0}^{-1/3} v_{\rm SW}^{-1/3}\Phi^{1/3}r^{2/3}.
\label{R}
\ea
\cite{BS1998} have found $R \propto r^{0.78\pm0.10}$, close to the value of  $2/3$ predicted by Equation (\ref{R}).

\section{Observational Signature}

\subsection{Magnetic Field Measured}
The \Bf\ measured by a magnetometer as a function of time depends on the relative velocity of the detector and spheromak boundary. 
Generally, the velocity of the  detector  with respect to the spheromak's center is dominated by the advection velocity of the spheromak with the Solar wind, which is typically a factor of several larger than velocity of expansion of the spheromak in the winds frame.

\begin{figure}[h!]
   \includegraphics[width=0.49\linewidth]{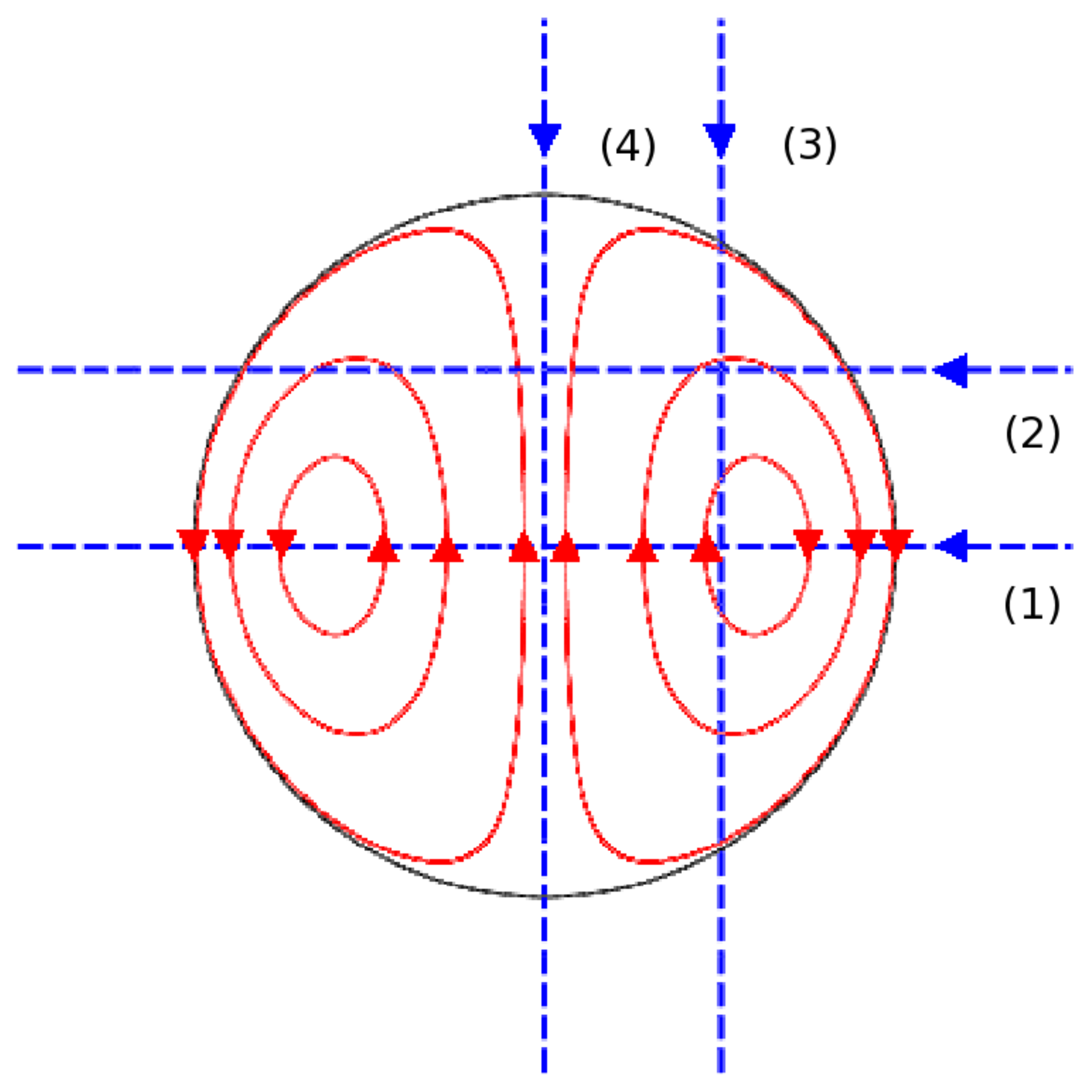}
      \includegraphics[width=0.49\linewidth]{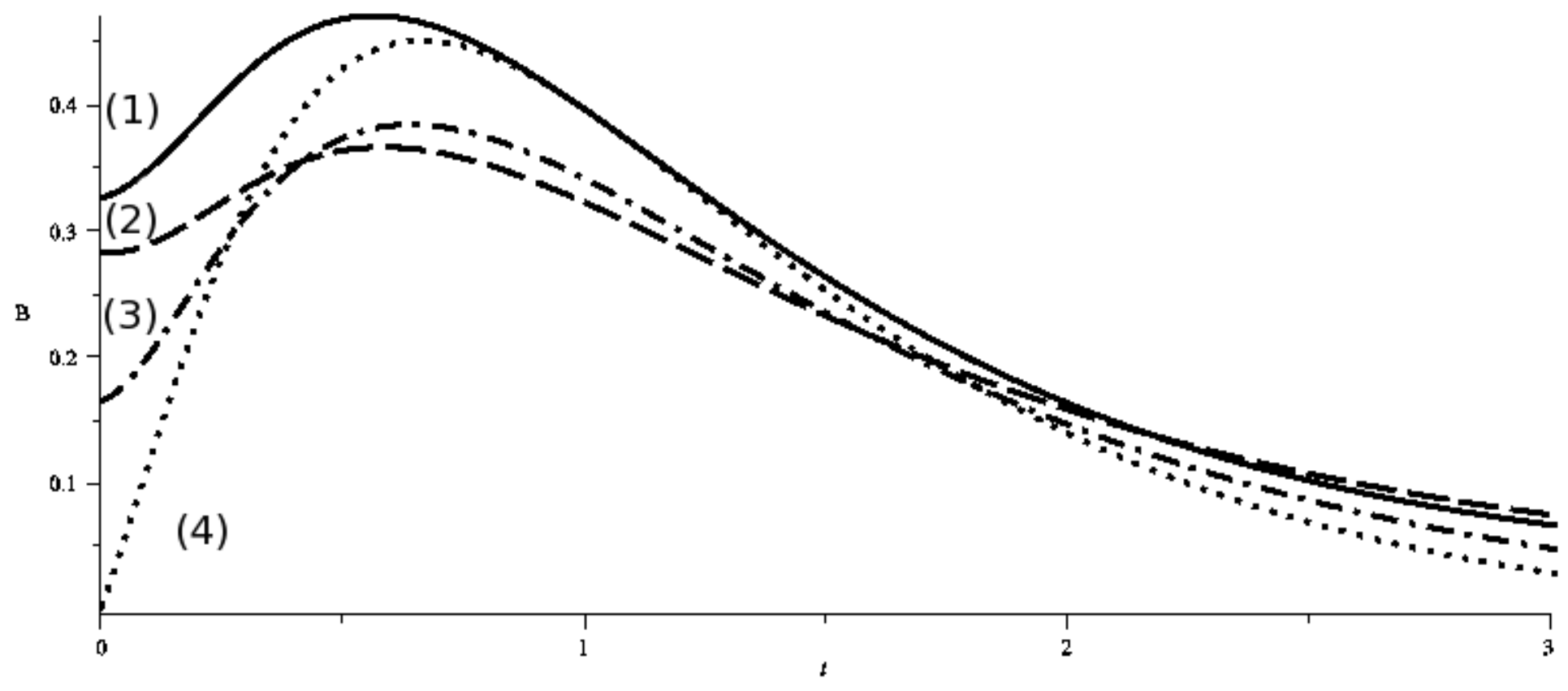}
   \caption{ Left Panel: Sections of a spheromak with a meridional plane. The trajectories of the detector crossing the spheromak are the dashed lines.  Right Panel:  Magnetic field measured by a detector flying through an expanding spheromak. The detector enters the cloud at $t=0$, at $t=1$ it crosses the axis: curves (1), (2); or the equator: curves (3), (4). We investigate four cases corresponding to the numbers of the trajectories shown in left panel. The solid curve (1) corresponds to a detector flying on the equator of the spheromak and passing from the centre at $t=1$. The dashed curve (2) corresponds to a detector that flies parallel to the equator and crosses the axis at $t=1$. The dashed-dotted (3) curve corresponds to a detector that flies parallel to the axis and crosses the equator at $t=1$. Finally the dotted curve (4) corresponds to a detector flying along the axis and passing from the centre at $t=1$. In all cases the maximum magnetic field is measured at $t<1$ before the detector crosses the middle of the cloud.  }
 \label{B_spheromak}
 \end{figure}

\begin{figure}[h!]
  \includegraphics[width=0.9\linewidth]{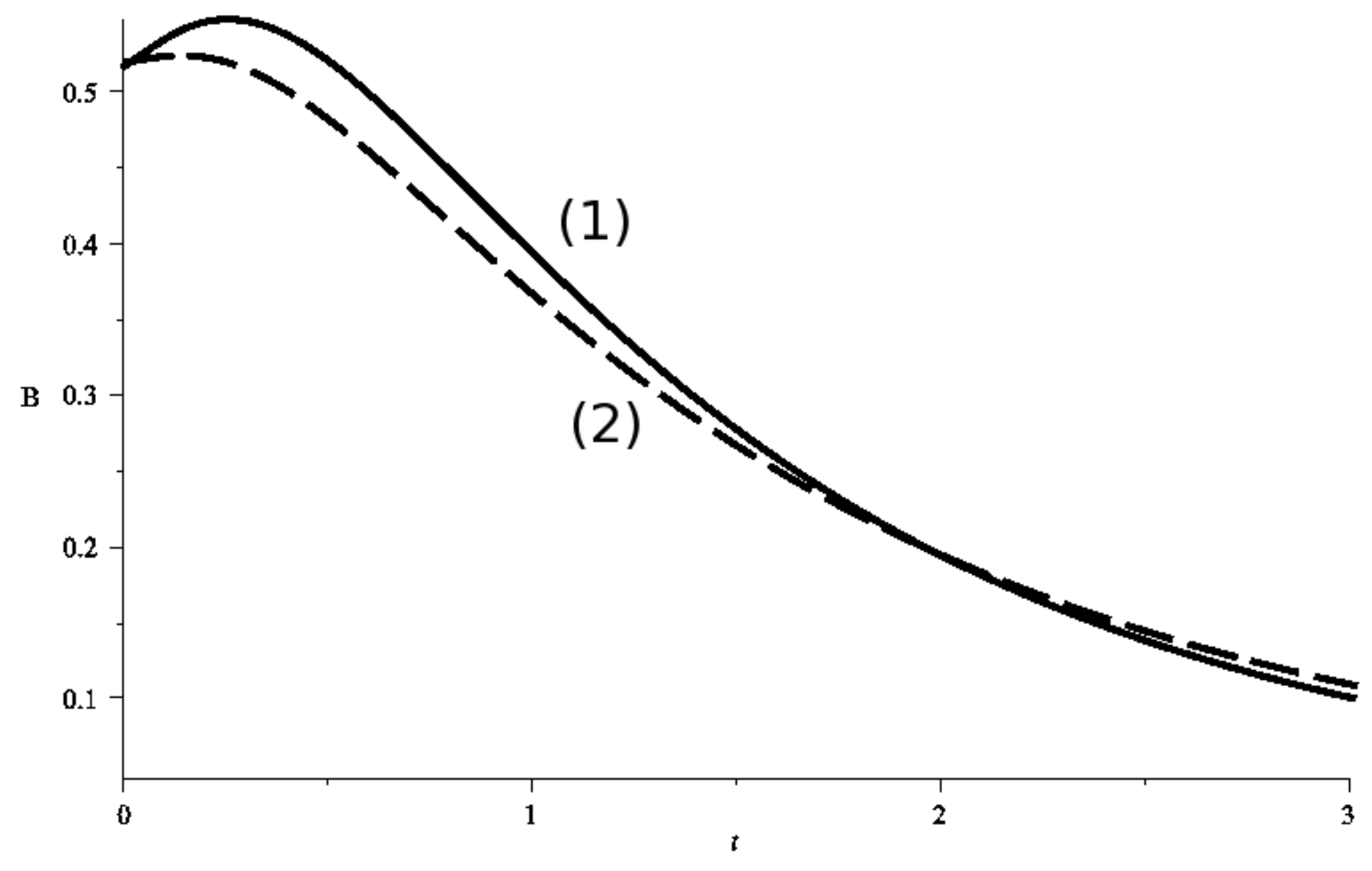}
  \caption{Magnetic field measured by a detector flying through an expanding cylinder. The solid curve (1) corresponds to a detector that flies and passes from the centre, whereas the dashed curve (2) correspond to a detector that enters the structure at an impact angle of $\pi/6$ and misses the centre by half radius. 
  }
\label{ropek0}
\end{figure}

Observations of 
 CMEs often indicate that  the maximum of B-field is reached before the spacecraft reaches the  geometrical center \citep{Burlaga,1993JGR....98.7621F}. It was long suspected that   is a consequence of the expansion of the magnetic cloud while it moves past the spacecraft  \citep[][Section 6.5.2]{Burlaga}:   when the detector reaches the middle of the spheromak, the  overall normalization
of the \Bf\ decrease due to expansion.
  Our model provides a quantitative description of this effect, see Figures \ref{B_spheromak} and \ref{ropek0}.
 
 \subsection{Plasma Velocity}

\begin{figure}[h!]
   \includegraphics[width=0.9\linewidth]{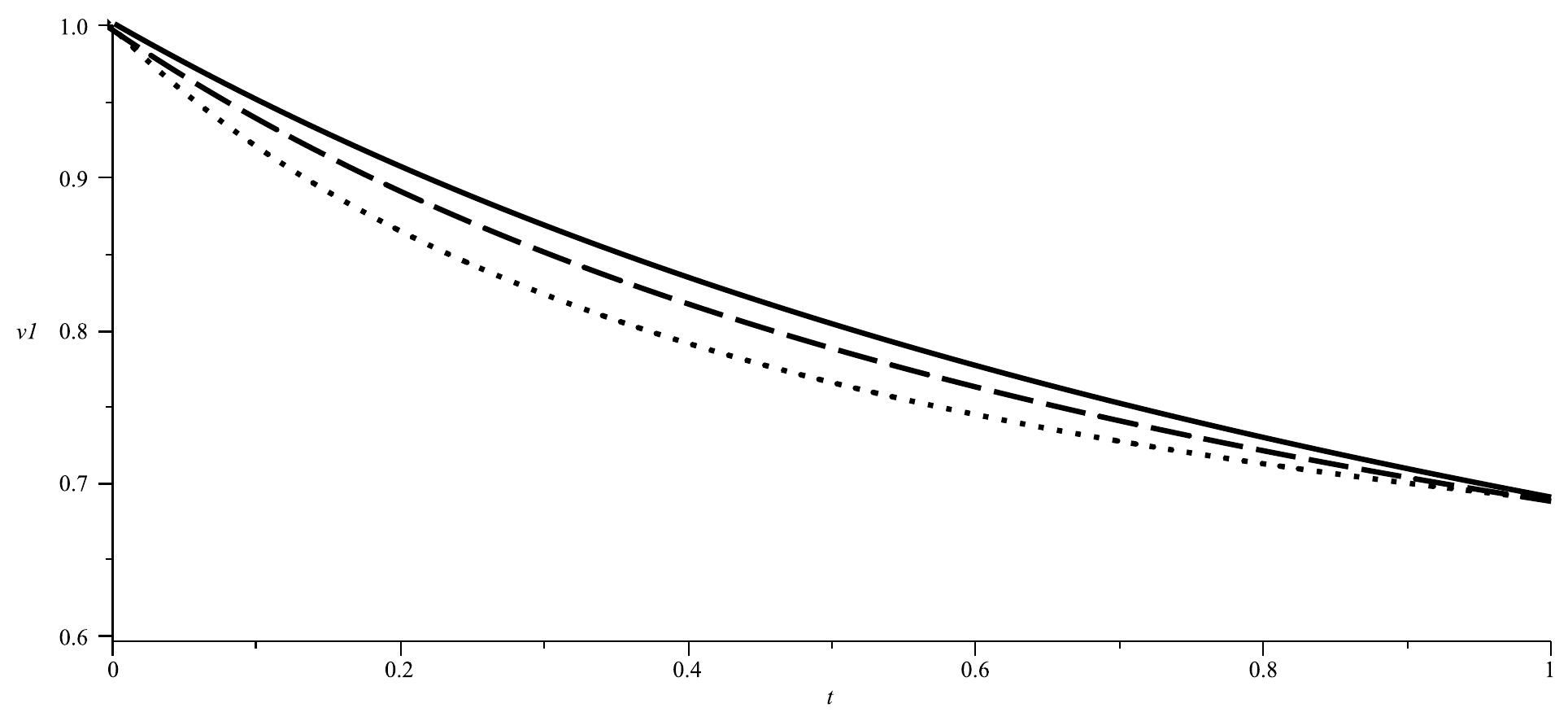} 
   \includegraphics[width=0.9\linewidth]{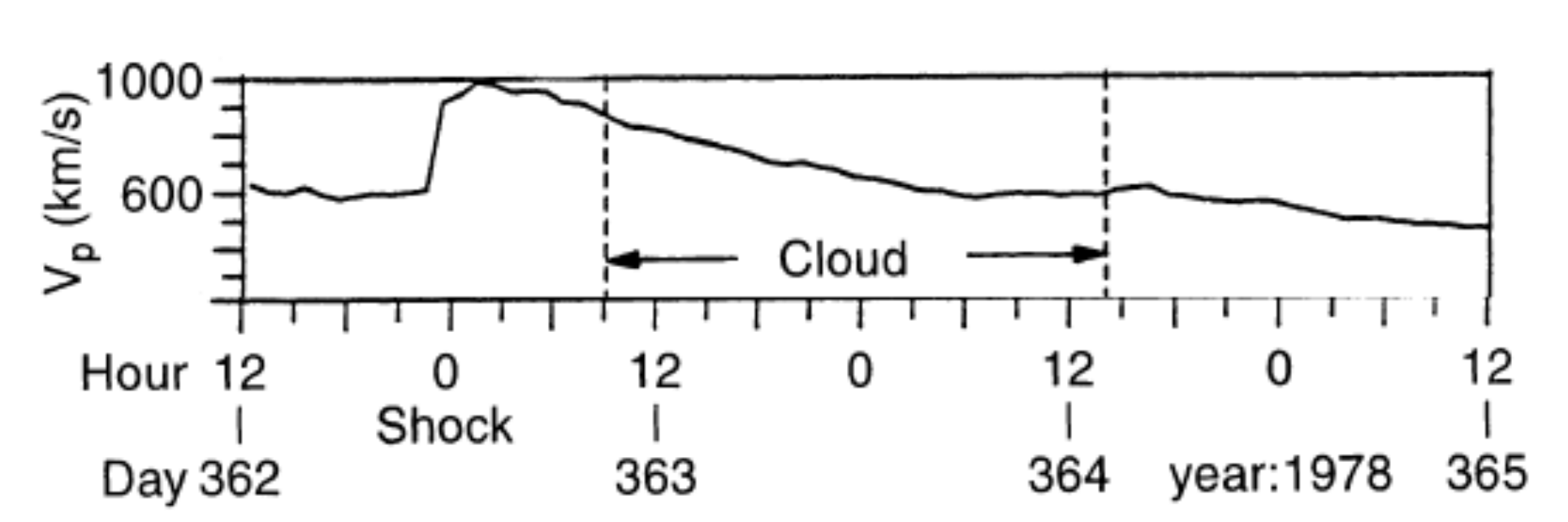}
   \caption{Upper panel: The velocity measured by a detector that encounters an expanding magnetic cloud and crosses its axis. These results are valid for both a spheromak cloud provided the detector's trajectory is on the equator and for a cylindrical one provided the trajectory is normal to the axis. We have assumed that the cloud propagates at constant velocity and expands self-similarly so that $R\propto t^{n}$. The detector measures the combined effect. The detector enters the cloud at $t=0$ and exits at $t=1$ where the velocity is $0.7$ of the initial. The solid line corresponds to a self-similar expansion with index $n=2/3$, the dashed to $n=1/2$ and the dotted to $n=1/3$.
   Lower panel: The velocity measured by the Helios of the event on days 362--365 1978 \citep{BS1998}.  For $n< 1$ the velocity profile in the cloud is consistent with the profile of the expanding model. }
 \label{Vel_M}
 \end{figure}

Observations suggest that the general trend of the velocity measured by a detector crossing the magnetic cloud, is to start from a large value and then decrease \citep{BS1998} to a smaller but comparable velocity. This is consistent with the image of a magnetic cloud that propagates from the Sun and at the same time expands. We have assumed that the propagation velocity is constant while the expansion is self-similar $R \propto t^{n}$, see Figure \ref{Vel_M}. In general, clouds differ in the details of the velocity profiles, but most of them give measurements that are in agreement with this model. 

\section{Discussion}

In this paper we discuss  the self-similar solutions for expanding force-free magnetic structures in spherical and cylindrical geometries. Under the assumption of non-relativistic MHD, when the dynamical effects of the resulting induced \Ef\ can be neglected, the structure remains force-free during expansion, ${\bf J } = \alpha {\bf B}$. In case of relativistic expansion, when \Ef\ has important dynamical effects \citep{Prendergast2005,GL2008},  spheromak-type self-similar solutions may be found for constant expansion velocity (R.~D.~Blandford, private communication).  In contrast, our solutions depend on $R$ and $\dot{R}$ measured at a given time, and thus are applicable to any $R(t)$.

In case of spheromaks, the internal velocity develops azimuthal component, $v_\phi \neq 0$, even though the surface expands purely radially. 
The fact that the \Bf\ structure remains force-free during expansion is qualitatively different from the
solutions of  \cite{Low82,Farrugia95}, which  were limited to pure radial motion. As a result,  even if  at the initial moment \Bf\ is force-free, during expansion non-zero pressure forces appeared.

We expect that the solution is dynamically and resistively stable, since 
stationary force-free solutions with constant $\alpha$  are very stable configurations \citep{Bellanbook}, both to ideal and  resistive instabilities, like tearing mode. 
Stability of expanding  force-free  structures should be investigated independently, but we expect them to be stable to global modes, since expansion-induced velocity shear will contribute to global stability (but may still result in development of local shearing instabilities).

Expansion of magnetic clouds is qualitative different from the 
self-similar  fluid  outflows, since in this case we have an additional constraint,  the conservation of magnetic helicity. Since the conservation of magnetic helicity and magnetic energy are, generally,  not compatible with each other, the self-similar expansion of spheromak can be connected only to 
 one  particular self-similar (Sedov-type) solution, with $r \propto t^{1/3}$. This solution has energy decreasing with time and thus requires strong energy losses. 

The model predicts that that the maximum \Bf\ is reached before the detector reaches geometrical center of a CME, while the plasma velocity measured by the spacecraft smoothly decreases with time. Both these facts are in general agreement with observations.

\section*{Acknowledgements}
We would like to thank Roger Blandford,  Charles Farrugia  and Martin Laming for insightful discussions.

 \bibliographystyle{plain}
\bibliography{BibTex.bib}

\end{document}